\documentclass[aps,pra,reprint,superscriptaddress]{revtex4-1}

\usepackage[dvips]{graphicx}
\usepackage{bm}
\usepackage{color}

\begin{document}


\title{{\color{black}Plasma concept for generating circularly polarized
  electromagnetic waves with relativistic amplitude}} 


\author{Takayoshi Sano}
\email{sano@ile.osaka-u.ac.jp}
\affiliation{Institute of Laser Engineering, Osaka University, Suita,
  Osaka 565-0871, Japan} 

\author{Yusuke Tatsumi}
\affiliation{Institute of Laser Engineering, Osaka University, Suita,
  Osaka 565-0871, Japan} 

\author{Masayasu Hata}
\affiliation{Institute of Laser Engineering, Osaka University, Suita,
  Osaka 565-0871, Japan} 

\author{Yasuhiko Sentoku}
\affiliation{Institute of Laser Engineering, Osaka University, Suita,
  Osaka 565-0871, Japan} 

\date{Nov 6, 2020; accepted for publication in Physical Review E}

\begin{abstract}
Propagation features of circularly polarized (CP) electromagnetic waves in magnetized plasmas are determined by the plasma density and the magnetic field strength. 
This property can be applied to design a unique plasma photonic device for intense short-pulse lasers. 
We have demonstrated by numerical simulations that a thin plasma foil
under an external magnetic field works as a polarizing plate to separate a linearly polarized laser into two CP waves traveling in the opposite direction.
This plasma photonic device has an advantage for generating intense CP waves even with a relativistic amplitude.
For various research purposes, intense CP lights are strongly required
to create high energy density plasmas in the laboratory.
\end{abstract}

\maketitle


\section{Introduction \label{sec1}}

Ultra-intense lasers are a unique tool to investigate extreme plasma
conditions, which brings new insight for various research fields such
as quantum beam science, laser fusion, and laboratory laser astrophysics. 
The essence of these applications is to control intense lasers to produce desired situations of the electromagnetic waves and plasmas for each purpose. 
Plasma photonic device emerges as an attractive medium for light manipulation at high intensities many orders of magnitude beyond the breaking point of traditional optics \cite{kodama04}. 
For example, plasma mirror works as an ultrafast optical shutter that enhances the pulse contrast \cite{kapteyn91}.  
Plasma grating is routinely used to deflect the flux of laser energy \cite{michel09,moody12}. 
Fruitful laser-plasma interactions open up the possibilities for various plasma devices such as pulse amplification \cite{malkin99,shi17} and higher harmonic production \cite{carman81}.

In this paper, we consider a plasma device to convert a linearly polarized (LP) laser to circularly polarized (CP) lights.
CP lasers with relativistic amplitude are thought to be an ideal means for laser-driven ion acceleration by radiation pressure \cite{esirkepov04} or collisionless shocks \cite{fiuza12} and effective ion heating by standing whistler waves \cite{sano19,sano20}.
However, it is difficult to increase the intensity of CP lasers
using traditional optics due to the fluence threshold with typically a
few J/cm$^2$. 

There are several ideas of plasma devices for CP-wave generation utilizing laser-plasma interactions, although there is a severe limit on the laser intensity. 
\citet{michel14,michel20} propose a novel photonic device as plasma waveplate by the use of optical two-wave mixing.
\citet{lehmann16,lehmann18} demonstrate numerically that transient plasma photonic crystal can transform LP laser light into circular polarization. 
It is known that relativistic transparency by anisotropic
electron distributions drastically alters the laser polarization
\cite{stark15}. 
Compared with these methods, the key difference of a scheme proposed
here is in the use of an external magnetic field.
Our plasma device has a quite simple design, but it can be applied to intense lasers exceeding the relativistic amplitude. 

The achievement of quasi-static strong magnetic fields generated in laser experiments shed light on the importance of the interactions between laser lights and magnetized plasmas \cite{fujioka13,santos18}.
Electromagnetic waves propagate in plasmas as whistler waves when the strength of an external magnetic field is sufficiently large.
The critical strength $B_c$ for the whistler wave is given by a condition that the wave frequency $\omega_0$ equals to the electron cyclotron frequency, i.e., $B_c = m_e \omega_0 / e$, where $m_e$ is the electron mass and $e$ is the elementary charge. 
Since there is no cutoff density for the whistler wave, it can
propagate in plasmas at any density {\color{black}\cite{budden61,yang15,luan16}}.
This essential feature stimulates magnetized plasma devices to manage the
transmission of CP waves \cite{ma16} and generate CP waves by the
Faraday effect \cite{weng17}.


We consider a plasma photonic device based on the unique characteristics of the whistler wave.
The outline of this paper is as follows.  
The design strategy for the generation of CP waves with relativistic amplitude is explained in Sec.~\ref{sec2}. 
Our interest is in CP waves transmitted after the interaction between a plasma foil and an LP laser along an external magnetic field.
Numerical setup and method of two-dimensional (2D) Particle-in-Cell (PIC) simulations are described in Sec.~\ref{sec3}. 
In Sec.~\ref{sec4}, a systematic survey of the simulations is performed for functional evaluation of the plasma device as a CP-wave filter.
The dependences of the transmittance and reflectivity on the magnetic field strength and the laser intensity are elucidated in detail.
We examine the effects of the angle between the laser and magnetic field directions in Sec.~\ref{sec5}.
The possibility of the L wave extraction is also discussed.
Finally, our conclusions are summarized in Sec.~\ref{sec6}.

\section{Underlying Concept \label{sec2}}

We consider a possible way to extract CP
waves from an LP laser by controlling laser-plasma
interactions under a uniform magnetic field.
When an external magnetic field is applied to a plasma target along
the direction of laser propagation, electromagnetic waves can travel
in plasma as either the R wave or the L wave. 
The R (L) wave is a right-hand (left-hand) CP light with respect to the magnetic field direction. 
The propagation features of these CP waves depend on the
electron density $n_e$ of the plasma and the magnetic field strength
$B_{\rm ext}$. 
Assuming the laser frequency is $\omega_0$, the refractive indexes for
the R and L waves are given by
\begin{eqnarray}
N_R^2 = 1 - \frac{\widetilde{n}_e}{1-\widetilde{B}_{\rm ext}} \;, \\
N_L^2 = 1 - \frac{\widetilde{n}_e}{1+\widetilde{B}_{\rm ext}} \;, 
\end{eqnarray}
where $\widetilde{n}_e = \omega_{pe}^2 / \omega_0^2 = n_e / n_c$ is the electron
density normalized by the critical density $n_c$, and
$\widetilde{B}_{\rm ext} = \omega_{ce} / \omega_0 =  B_{\rm ext} / B_c$ is the ratio of the external magnetic field to the critical value $B_c$.
Here, $\omega_{pe}$ and $\omega_{ce}$ are the electron plasma frequency
and cyclotron frequency, respectively.


\begin{figure*}
\includegraphics[scale=0.85,clip]{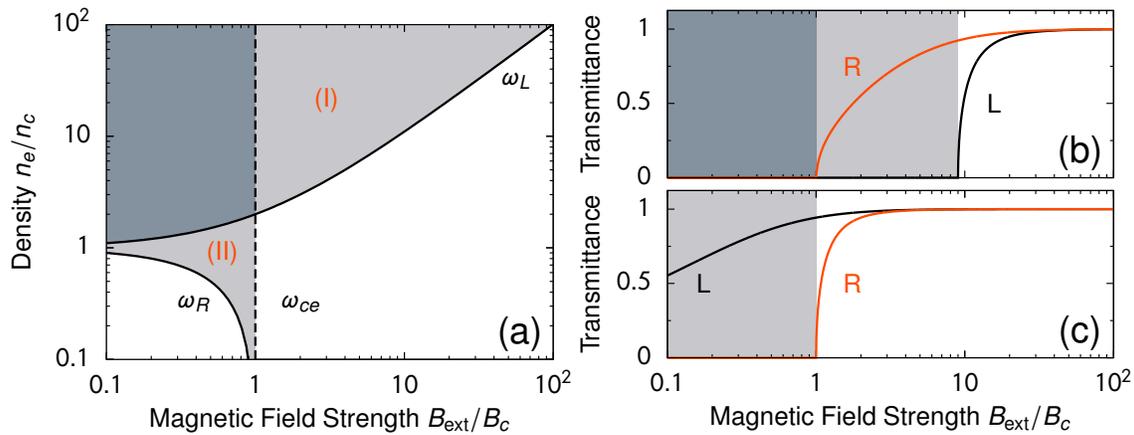}%
\caption{
(a)
Propagation properties of the electromagnetic waves in plasmas along an
external magnetic field.
The allowance of the propagation for the R and L waves depends on the
plasma density $n_e/n_c$ and the magnetic field strength $B_{\rm
  ext}/B_c$.
The region (I) stands for where the R wave propagates but the L
wave is forbidden, and the region (II) is the opposite.
Both waves are allowed to propagate in the white regions, whereas
both are forbidden in the dark-gray region.
The boundary curves of the separated regions are the cutoff conditions
($\omega_R = \omega_0$ and $\omega_L = \omega_0$) and the resonance
condition ($\omega_{ce} = \omega_0$).
(b,c) Transmittance of the R and L waves, $T_R$ (red) and $T_L$
(black), as a
function of the magnetic field strength for the cases of (b) $n_e/n_c
= 10$ and (c) $n_e/n_c = 1$. 
\label{fig1}}
\end{figure*}

We focus on the transmittance of the R and L
waves at the interface between the vacuum and a plasma target.
Since the electromagnetic waves can propagate in plasmas only when 
$N^2$ is positive in the linear amplitude limit.
Figure~\ref{fig1} indicates the propagation features along a magnetic
field summarized in the density and magnetic field diagram ($\widetilde{n}_e$
vs. $\widetilde{B}_{\rm ext}$), which is the so-called  
CMA (Clemmow-Mullaly-Allis) diagram \cite{stix92}.
Here, the cold plasma approximation is adopted, and the ion mass is
assumed to be much higher than the electron mass. 
The parameter space is divided into several regions by the cutoff
conditions for the R and L waves ($\omega_R = \omega_0$ and $\omega_L
= \omega_0$) and the cyclotron resonance condition ($\omega_{ce} =
\omega_0$), where the cutoff frequency for each wave is written as
\begin{eqnarray}
\frac{\omega_R}{\omega_0} = \frac12
\left[ \widetilde{B}_{\rm ext} + \left( 
\widetilde{B}_{\rm ext}^2 + 4 \widetilde{n}_{e} \right)^{1/2} 
\right] \;, \\
\frac{\omega_L}{\omega_0} = \frac12 
\left[ - \widetilde{B}_{\rm ext} + \left( 
\widetilde{B}_{\rm ext}^2 + 4 \widetilde{n}_{e} \right)^{1/2} 
\right] \;.
\end{eqnarray}
This diversity in the propagation features is quite different from the normal cutoff condition
($\omega_{pe} = \omega_0$ or $\widetilde{n}_e = 1$) without the
external magnetic field.

The forbidden regions for the R and L wave are different.
An LP light is regarded as a combined wave of a right-hand CP (RCP)
and a left-hand CP (LCP) wave. 
If one of the CP waves is allowed to enter the plasma target, but the
other is forbidden, the plasma target will work as a filter to
generate a CP wave from an LP light. 
Two regions satisfy such conditions, which are labeled by (I) and (II) in
Fig.~\ref{fig1}(a).

When the magnetic field strength is larger than the critical $B_c$,
the whistler-wave mode of the R wave propagates in plasmas at any
density.
On the other hand, there is a cutoff density for the L-wave component.
Only the R wave is allowed to pass through the plasma target in
the region (I). 
Since the L-cutoff density depends on the field strength as $\widetilde{n}_L
= \widetilde{B}_{\rm ext} + 1$, the parameter range of the region (I) is
given by
\begin{equation}
1 < \widetilde{B}_{\rm ext} < \widetilde{n}_e - 1 \;.
\end{equation}
The transmittance of the R and L wave at the vacuum-plasma boundary is
evaluated by the refractive indexes according to the Fresnel formula
\cite{budden61}. 
The energy fraction of the transmitted wave relative to the injected
energy of each CP wave is derived as $T_{\ast} = 4 N_{\ast} / (1 +
N_{\ast})^2$, where the subscript $\ast$ denotes $R$ or $L$.
The reflectivity is then $R_{\ast} = (1 - N_{\ast})^2 / (1 + N_{\ast})^2$.

Figure~\ref{fig1}(b) shows the transmittance $T_R$ and $T_L$ for
a plasma density with $\widetilde{n}_e = 10$.
The region (I) corresponds that $1 < \widetilde{B}_{\rm ext} < 9$ for
this case, which is highlighted by light-gray in the figure.
The most ideal situation for the R-wave generation is $T_R = 1$ and $T_L
= 0$.
In the region (I), the L wave is perfectly reflected, whereas the
transmittance for the R wave increases with the magnetic field
strength. 
Then the highest transmittance for the whistler wave will be realized
at just below the L-cutoff, that is $\widetilde{B}_{\rm ext} \sim
\widetilde{n}_e - 1$.
This magnetic field strength is the optimized condition by which
the plasma foil target will work as an R-wave filter most efficiently.
The higher transmittance could be realized for much denser targets as
well, although the optimal strength of the magnetic field becomes 
larger $\widetilde{B}_{\rm ext} \sim \widetilde{n}_e \gg 1$.

By contrast, only the L wave is allowed to propagate in the region
(II) of Fig.~\ref{fig1}(a). 
The parameter range of the region (II) can be expressed as
\begin{equation}
| \widetilde{n}_e - 1 | < \widetilde{B}_{\rm ext} < 1 \;.
\end{equation}
The required strength of the magnetic field is much smaller compared to
that of the region (I), which could be an advantage in terms of
practical application.
The transmittance for the case of $\widetilde{n}_e = 1$ is depicted
in Fig.~\ref{fig1}(c).
The region (II) for this case is the range of $\widetilde{B}_{\rm ext} < 1$.
The transmittance of the L wave increases with
$\widetilde{B}_{\rm ext}$, whereas the R wave starts to enter the plasma
when $\widetilde{B}_{\rm ext} > 1$.
Then the best performance as an L-wave filter ($T_R = 0$ and $T_L \sim
1$) will be achieved at just
below the resonance condition, that is $\widetilde{B}_{\rm ext} \sim 1$.
Because of the lower density of the region (II), the
relativistic effects may not be ignored when the incident laser
amplitude is relativistic (see more Sec. \ref{sec5}).

By controlling such propagation properties, we can
design a plasma device that converts an LP light into CP waves.
In this paper, we concentrate our discussion on the region (I) where
only the whistler wave gets into the plasma.
In the following, we will search for the optimal conditions of the
plasma photonic device to generate CP lights with the help of
multi-dimensional PIC simulations. 

\section{Numerical Setup \label{sec3}}

The interaction between a short-pulse laser and a dense plasma foil
is examined in our numerical experiment. 
We consider a simple situation in which a thin foil target of solid
hydrogen is irradiated by a Gaussian-shape laser. 
The plasma target is assumed to be fully ionized initially.
The number densities of the electrons and ions, $n_e = n_i$, are
spatially constant inside the target. 
For simplicity, the foil is in contact directly with the vacuum
without any preplasma.  
The thickness of the target $L$ is set to be the same as the laser
wavelength $\lambda_0$ in the vacuum. 
Hereafter the length scale is normalized by the laser wavelength such
as $\widetilde{L} = L / \lambda_0 = 1$.

The injected laser is an LP light traveling in the $x$
direction. 
The pulse duration at half maximum amplitude is $\widetilde{\tau}_0 =
\tau_0 / t_0 = 3$, where the unit of time scale is given by the
laser period $t_0 = \lambda_0 / c$ and $c$ is the speed of light. 
The incident angle of the laser is normal to the target surface, and 
the focal spot is $\widetilde{W} = 3$ in diameter.
The peak laser intensity $I_0$ is parameterized by the normalized vector
potential $a_0 = e E_0 / (m_e c \omega_0)$ via the relation of $I_0 =
\epsilon_0 c E_0^2 / 2$, where $E_0$ is the laser electric field and
$\epsilon_0$ is the vacuum permittivity. 
A uniform external magnetic field $B_{\rm ext}$ is applied parallel to the
laser propagation direction.  

Numerical simulations on the laser-plasma interaction in Cartesian
coordinates ($x$, $y$) are calculated by a PIC scheme, PICLS
\cite{sentoku08}.
The collision term is ignored in this analysis, unless otherwise mentioned.
The computational domain is set to be $(\widetilde{D}_x , \widetilde{D}_y)
= (13, 9)$, and the plasma target is located at the middle $-0.5 \le
\widetilde{x} \le 0.5$.
The escape boundary conditions for the electromagnetic waves and
plasma particles are applied at both side of the $x$ boundaries,
whereas the periodic boundary conditions are adopted in the $y$
direction. 
The spatial and time resolutions are $\Delta \widetilde{x} = \Delta
\widetilde{t} = 5 \times 10^{-3}$ and the particle
number is 100 per each grid at the beginning.
The third order interpolation algorithm is used to suppress the
numerical heating.
The time resolution $\Delta t$ should
be sufficiently smaller than the electron gyration time as well as the
plasma oscillation time to avoid the unphysical numerical heating in
strongly magnetized plasmas.  

The transmittance and reflectivity of the electromagnetic waves are
assessed by integrating the energy of the waves passing
through each side of the $x$ boundary. 
The calculations are continued long enough until the transmittance
and reflectivity become constant with time, which is typically
$\widetilde{t}_{\rm end} = 25$ after the injected laser hits the target.
The key parameters of this system are the target density $\widetilde{n}_e$, the
magnetic field strength $\widetilde{B}_{\rm ext}$, and the laser
amplitude $a_0$. 
The fiducial values representing the region (I) 
are selected as $\widetilde{n}_e = 10$, $\widetilde{B}_{\rm ext} = 7$,
and $a_0 = 1$.
The dimensional quantities of these fiducial parameters are estimated as a
function of the laser wavelength $\lambda_0$.
For instance, the laser intensity is given by 
$I_0 = 1.4 \times 10^{18} a_0^2 \; \lambda_{{\mu}{\rm m}}^{-2}$ 
W/cm$^2$, where $\lambda_{{\mu}{\rm m}}$ is the laser wavelength in microns.
The critical density is calculated as
$n_c = 1.1 \times 10^{27} \; \lambda_{{\mu}{\rm m}}^{-2}$ m$^{-3}$,
and the critical magnetic field strength is 
$B_{c} = 11 \; \lambda_{{\mu}{\rm m}}^{-1}$ kT.
For this case, the unit time corresponds to 
$t_0 = 3.3 \; \lambda_{{\mu}{\rm m}}$ fs. 

\section{Numerical Results \label{sec4}}

The transmittance and reflectivity of the injected LP
laser light are evaluated by 2D PIC simulations.
In this section, the density of the plasma target is fixed to be
$\widetilde{n}_e = 10$, and then we will seek conditions that only
the whistler wave is selectively transmitted. 
The foil thickness is $\widetilde{L} = 1$, and the amplitude and pulse
duration of the laser are $a_0$ = 1 and $\widetilde{\tau}_0 = 3$.
After enough time has passed since the laser-plasma interaction, the
incident laser energy should be transferred to four different forms:
the transmitted and reflected electromagnetic-wave energies and the
absorbed ion and electron kinetic energies.


\begin{figure}
\includegraphics[scale=0.85,clip]{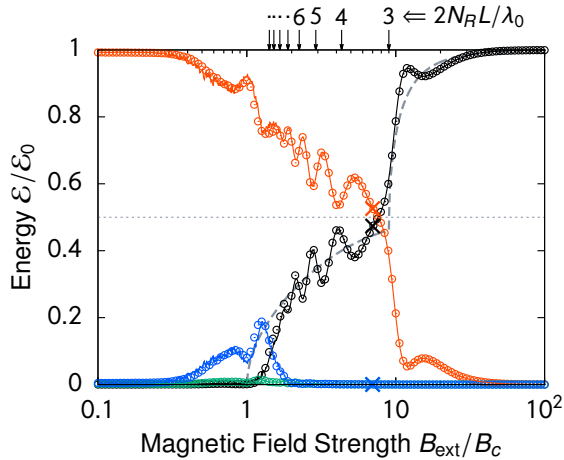}%
\caption{
Transmittance (black) and reflectivity (red) after the interaction
between laser and magnetized plasma target evaluated numerically by a
series of PIC simulations.
The kinetic energies of ions (green) and electrons (blue) at the end of
each run are also indicated. 
The energy fractions are normalized by the incident laser energy
${\cal{E}}_0$. 
The plasma density is assumed to be $n_e / n_c = 10$ and the laser
amplitude is $a_0 = 1$. 
The circles are obtained from 2D runs (120 runs).
The same calculations performed in the 1D and 3D geometries are also shown
by the thin solid curves (600 runs) and cross marks (1 run), respectively.
The gray dashed curve is the theoretical prediction of the
transmittance.
{\color{black}
The interference condition $2 N_R L / \lambda_0$ is indicated at the top
of the figure, where $N_R$ has a dependence of $B_{\rm ext}$.}
\label{fig2}}
\end{figure}

The outcome of the energy redistribution after the laser-plasma
interaction depends on the strength of the external magnetic field.
The transmittance and reflectivity in terms of the incident laser
energy are depicted by the black and red circles in Fig.~\ref{fig2}.
The final state of the ion and electron kinetic energies are also
shown by the green and blue circles, respectively.
The energy conversion to plasmas is not significant
except for the cases around $\widetilde{B}_{\rm ext} \sim 1$, where the
electrons absorb the wave energy through the cyclotron resonance
\cite{sano17}.
Then, most of the laser energy remains in the form of electromagnetic waves. 

\begin{figure}
\includegraphics[scale=0.85,clip]{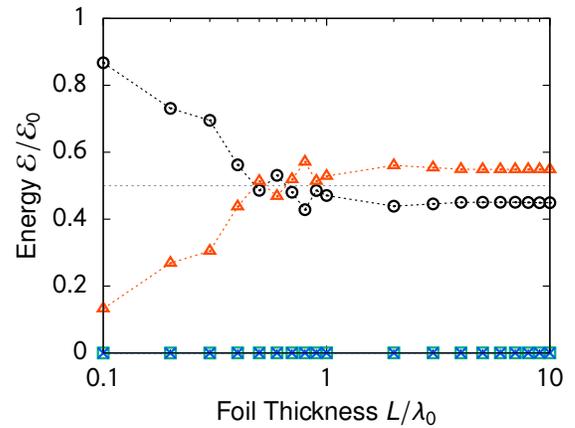}%
\caption{
Dependence of the transmittance (black circles) and refectivity (red
triangles) on the thickness of
target foil $L$ obtained by 1D PIC simulations. 
The kinetic energy of ions (green squares) and electrons (blue
crosses) are also shown.
The parameters except for the foil thickness are the density $n_e /
n_c = 10$, the magnetic field strength $B_{\rm ext} / B_c = 7$, and
the laser amplitude $a_0 = 1$. 
The transmittance is independent of the foil thickness when $L /
\lambda_0 \gtrsim 0.5$. 
\label{fig7}}
\end{figure}

The numerically obtained transmittance is consistent with the
theoretical estimation shown by the gray dashed curve in Fig.~\ref{fig2}.
Because an LP light is a sum of an RCP light and an LCP light, each
having a half component, the total transmittance at a given 
$\widetilde{B}_{\rm ext}$ is basically the average of $T_R$ and $T_L$ shown in
Fig.~\ref{fig1}(b). 
Note that the effect of multiple reflections on both sides of the target surface is taken into account in the theoretical curve.
The numerical results of the transmittance seem to oscillate around
the predicted curve in the range of 
$1 \lesssim \widetilde{B}_{\rm ext} \lesssim 10$. 
The oscillation is due to the interference by multiple reflections
inside the target. 
The oscillation peaks in the transmittance appear when an integer
multiple of the whistler wavelength {\color{black}($\lambda_0 / N_R$)}
coincides with 
the optical path difference of $2 L$ {\color{black}(see the labels at
  the top of Fig. 2)}. 
When the target thickness becomes much thicker than the whistler
wavelength (e.g., $\widetilde{L} \gtrsim 3$), the oscillatory behavior
disappears, and the numerical results exhibit an excellent agreement
with the theory. 
If the target is sufficiently thicker than the electron skin depth $c
/ \omega_{pe}$, the transmittance feature is independent of $L$ (see
Fig.~\ref{fig7}). 


\begin{figure*}
\includegraphics[scale=0.85,clip]{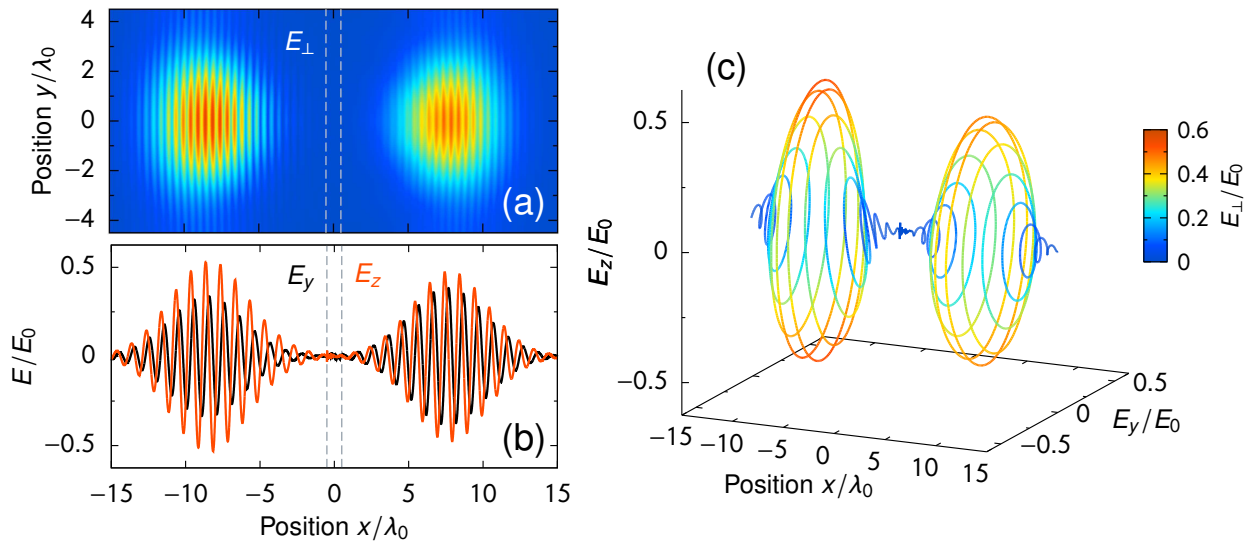}%
\caption{
(a)
Spatial distribution of the perpendicular
electric field $E_{\perp} = (E_y^2 + E_z^2)^{1/2}$ of the transmitted
($x > 0$) and reflected ($x < 0$) waves, where the plasma foil is
located at $-0.5 \le x / \lambda_0 \le 0.5$ (gray dashed line). 
In this 2D fiducial run, the plasma density is $n_e / n_c =
10$, the magnetic field strength is $B_{\rm ext} / B_c = 7$, and the laser
amplitude is $a_0 = 1$.
The pulse duration of the incident laser is $\tau_0 / t_0 = 3$ and the
focal spot size at the foil surface is $W / \lambda_0 = 3$ in the
$y$-direction. 
{\color{black}
This snapshot is taken at a time $t / t_0 = 11.2$ after the incident
laser reaches the target surface.}
Only for this run, a wider domain size in the $x$ direction is used
($D_x / \lambda_0 = 31$).
(b)
1D distributions of the electric fields, $E_y$ (black) and $E_z$ (red),
along the $x$-axis extracted from the same snapshot shown in (a). 
(c)
3D plot of the direction of $E_{\perp}$ using the same data as in (b).
The color denotes the amplitude of the perpendicular electric field. 
\label{fig3}}
\end{figure*}

In this analysis, we look for the physical parameters of a plasma
device that is transparent for the R wave and entirely reflective for
the L wave. 
It is confirmed that the ideal situation as for a CP wave generator
is realized when the magnetic field strength is just below the
L-cutoff condition, that is 
$\widetilde{B}_{\rm ext} \sim 9$.
Figure~\ref{fig3} shows snapshot images of the electromagnetic
waves in one of the best parameters as the plasma device.
The field strength of this run is $\widetilde{B}_{\rm ext} = 7$.

The spatial distribution of the perpendicular component of the
electric field $E_{\perp}$ is indicated in Fig.~\ref{fig3}(a).
The target foil is located at the center, and the incident laser comes
from the left.
The energy conversion efficiency from the injected LP
light to the transmitted wave is 0.47.
At the time of the snapshot, the transmitted wave and the reflected
wave exist almost symmetrically to the target.  
Figure~\ref{fig3}(b) is a 1D distribution of the electric
field along the $x$-axis.
The phase delay of $\pi/2$ between $E_y$ and $E_z$ indicates that the
transmitted wave is the R wave and the reflected wave is the L wave.
The amplitude of the transmitted and reflected wave is about half of the
incident laser amplitude, $a_T \sim a_R \sim a_0 / 2$.
{\color{black}
The polarization is mostly circular, but slightly elliptical.
For example, the deviation of the amplitude between $E_y$ and $E_z$ is
around 20\% in the transmitted wave, which might be due to the
contamination of components other than the whistler-wave mode.}
Figure~\ref{fig3}(c) is a 3D plot of the direction of the
$\bm{E}_{\perp}$ vector shown in Fig.~\ref{fig3}(b). 
It is evident that two CP waves are generated successfully after the
laser-plasma interaction. 

Note that both of the transmitted and reflected CP waves are RCP waves with
respect to the propagation direction.
The magnetic field direction is parallel to the laser injection.
The transmitted wave is the R wave so that the circular polarization
is right-hand to the propagation direction. 
The reflected wave, on the other hand, is the L wave, and then it has
left-hand polarity to the magnetic field direction.
Because the propagation direction is opposite to the magnetic field,
the reflected wave is also right-hand to the propagation direction.
Then, this plasma device generates two RCP
waves propagating in the opposite direction.
It might be possible to establish an optics system for combining these
two CP lasers in practical experiments. 

Interestingly, if the magnetic field direction is anti-parallel to the
laser propagation (i.e., $\widetilde{B}_{\rm ext} = - 7$), two
LCP waves will be formed by the same process. 
Therefore, the polarity of the generated CP
waves can be selected by the direction of the external magnetic field.

The laser-plasma interaction considered here is independent of the
spatial dimensions of the simulations.
In Fig.~\ref{fig2}, the 1D and 3D PIC results are plotted by the solid curves and cross marks, respectively. 
As seen from the figure, all the data exhibit the same trend, and which
indicates that the effect of the finite laser spot must be negligible.
The numerical resolution of the 1D runs is identical to that of the
2D simulations.
The conclusions illustrated by Fig.~\ref{fig2} are unaffected by the
numerical resolutions, which has been tested by the higher resolution
runs in 1D with $\Delta \widetilde{x} = 10^{-3}$.
It is also verified by 1D runs that the inclusion of the collisional
effects has little influence on the transmittance in our system. 
The fiducial run is also performed in 3D, where the grid resolution is
$\Delta \widetilde{x} = 10^{-2}$ and fewer particle numbers (20 per
grid) are solved. 
These numerical results demonstrate that the discovered feature of
this plasma device is robust and quite practical.

\begin{figure}
\includegraphics[scale=0.85,clip]{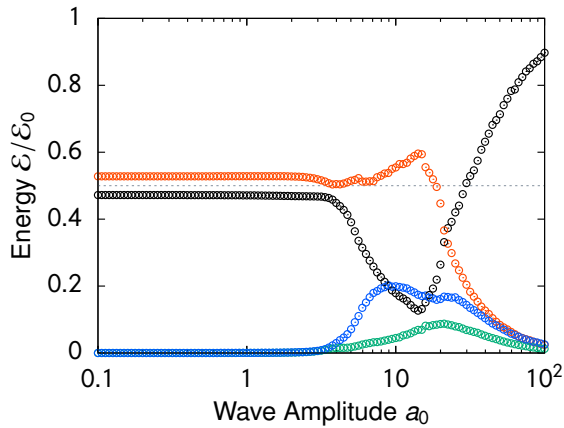}%
\caption{
Dependence of the transmittance (black) and reflectivity (red) on the
laser intensity.
The kinetic energies absorbed by ions (green) and electrons (blue) are
also shown.  
The energy fractions obtained from 2D PIC simulations are given by the
ratio to the incident laser energy ${\cal{E}}_0$. 
The fiducial parameters are used for the plasma density $n_e / n_c =
10$ and the magnetic field strength $B_{\rm ext} / B_c = 7$.
\label{fig4}}
\end{figure}

The most significant advantage of this plasma device is that it is
applicable even for the relativistic amplitude of the wave.
If the amplitude of the injected wave is relativistic, the
transmitted wave amplitude could also be nearly relativistic.
Figure~\ref{fig4} denotes the dependence of the transmittance on the
laser amplitude $a_0$.
The fiducial parameters, $\widetilde{n}_e = 10$ and $\widetilde{B}_{\rm
  ext} = 7$, are adopted for these 2D simulations.
When $a_0 \gtrsim 10$, some particles have escaped from the
computational domain, but those are not counted in the energies of
this figure.
The escaped energy is at most 10\% of the injected laser energy.


The transmittance is independent of the laser intensity until $a_0
\lesssim 3$.
The polarization property of the transmitted wave is also unchanged
for this range of $a_0$.
These facts confirm the applicability of this mechanism in the
relativistic intensity regime. 
It should be noticed that even when the amplitude of the transmitted
whistler wave is relativistic, $a_T = 2 a_0 / (N_R + 1) > 1$,  the
electron quiver velocity remains still non-relativistic, $v_q / c  =
a_T / (\widetilde{B}_{\rm  ext} - 1)$.
It is because the magnetic field considered here is sufficiently large
enough. 

If the laser amplitude becomes further higher, there are two
relativistic effects that prevent the CP wave generation.
One is the electron heating through the cyclotron resonance.
The external magnetic field in these runs is far beyond the resonance
condition.
However, due to the relativistic correction, the resonance
condition becomes $\omega_0 \sim \omega_{ce}/\gamma$, where $\gamma$
is the Lorentz factor of electrons.
The electrons in the target gain the energy efficiently through the resonance
when $a_0 \gtrsim (\widetilde{B}_{\rm ext}^2 - 1)^{1/2} \sim 7$
\cite{sano17}. 
In fact, the electromagnetic energy entered into the target is
largely absorbed by the electrons, and the transmittance decreases
drastically in the range of $7 \lesssim a_0 \lesssim 10$. 

The sudden rise in the transmittance at $a_0 \sim 10$ is caused by
the other effect, that is the relativistic transparency
\cite{kaw70,sakagami96}.  
The cutoff density for the relativistic-amplitude laser is
approximately given by $\gamma n_c$.
The minimum intensity for the relativistic transparency is obtained by
$a_0 \gtrsim [ 2 ( \widetilde{n}_e^2 - 1 ) ]^{1/2} \sim 14$, where the
quiver energy $\gamma = (1 + a_0^2/2)^{1/2}$ is assumed.
This is also consistent with the numerical results shown in
Fig.~\ref{fig4}. 
Therefore, the maximum intensity valid for the CP wave generator could be
managed by the field strength and the target density.

\section{Discussion \label{sec5}}

For practical use of this plasma photonic device, it might be difficult to
align the magnetic field direction precisely the same as the laser
direction. 
Then, it would be important to examine the angle dependence of
the transmittance.
The laser injection is kept to be along the $x$-axis in the following analysis.
We define $\varphi$ as an angle between the laser and magnetic
field directions in the unit of degree.
Here the angle is assumed to vary within the $x$-$y$ plane.
The angle $\varphi = 90$ (0) denotes that the magnetic field
direction is along the $y$-axis ($x$-axis).

2D PIC simulations reveal the angle dependence of the efficiency of our
plasma device.
The angle dependence of the phase velocity in the region (I)
is characterized by a figure-eight shape of the wave normal surface
\cite{stix92}. 
Thus, the transmittance of the R-wave branch should have a dependence
on the angle $\varphi$.
The polarization angle of the incident LP wave is another
independent parameter, so that we examine two cases, ${\bm{E}}_0
\parallel {\bm{e}}_y$ and ${\bm{E}}_0 \parallel {\bm{e}}_z$.
In the former cases, the electric field of the incident laser has a
parallel component to the external magnetic field.
However, the laser electric field is always perpendicular to $B_{\rm
  ext}$ in the latter cases.


\begin{figure*}
\includegraphics[scale=0.85,clip]{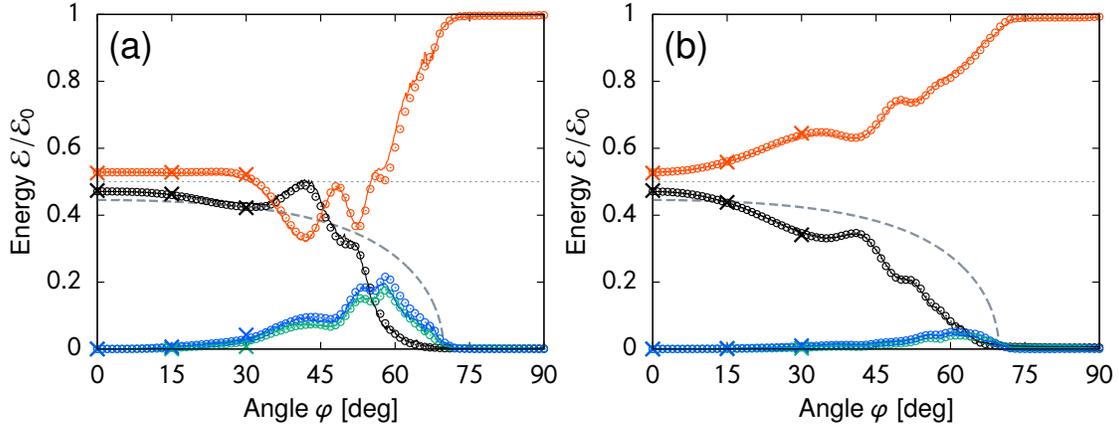}%
\caption{
Angle dependence of the transmittance (black), reflectivity (red), and
the energy fraction of ions (green) and electrons (blue).
The definition of $\varphi$ is the angle between the laser
direction and the magnetic field orientation. 
The fiducial parameters are used for the plasma density $n_e / n_c =
10$, the magnetic field strength $B_{\rm ext} / B_c = 7$, and the laser
amplitude $a_0 = 1$.
The laser-plasma interaction is affected by the electric field
direction of the incident LP laser.
The results when the laser electric field is in the 
$y$-direction and $z$-direction are shown in (a) and (b), respectively.
The indicated marks are the same as in Fig.~\ref{fig2}.
\label{fig5}}
\end{figure*}


\begin{figure*}
\includegraphics[scale=0.85,clip]{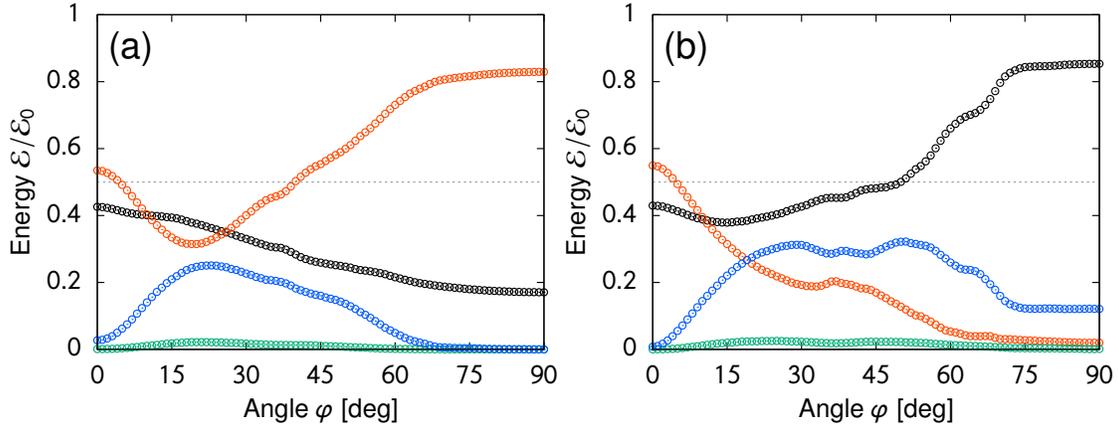}%
\caption{
Angle dependence of the transmittance (black), reflectivity (red), and
the energy fraction of ions (green) and electrons (blue).
The parameters representing the region (II) in Fig.~\ref{fig1}(a) are
chosen to be the plasma density $n_e / n_c =
1$, the magnetic field strength $B_{\rm ext} / B_c = 0.7$, and the laser
amplitude $a_0 = 0.1$.
The results when the laser electric field is in the 
$y$-direction and $z$-direction are shown in (a) and (b), respectively.
The indicated marks are the same as in Fig.~\ref{fig3}.
\label{fig6}}
\end{figure*}
          
Figures~\ref{fig5}(a) and \ref{fig5}(b) show
the angle dependence of the energy redistribution for the cases that 
the initial electric field is parallel to the $y$-axis and $z$-axis,
respectively. 
Except for the angle $\varphi$, the model parameters are identical to
the fiducial run shown by Fig.~\ref{fig3}, in which the plasma
density is $\widetilde{n}_e = 10$, the magnetic field strength is
$\widetilde{B}_{\rm ext} = 7$, and the incident laser amplitude is
$a_0 = 1$.
The fiducial run corresponds to the angle $\varphi = 0$ case.

It is found that the transmittance feature is roughly independent of
the angle $\varphi$ when the angle is less than about 15 degrees.
These results suggest the angle tolerance for the installation of this
plasma device. 
The transmittance of the R-wave branch decreases with the increase of the
angle. 
When the angle is $\varphi = 90$, the injected wave is totally
reflected. 
As a reference, the theoretical curve is indicated by 
the gray dashed curve in Fig.~\ref{fig5}, which corresponds to half of
the transmittance for the R-wave branch.
There is a critical angle $\varphi_c = 69.9$ where the transmittance,
or the phase velocity of the electromagnetic wave, becomes zero
\cite{stix92}. 
When the electric field has the $y$ component [Fig.~\ref{fig5}(a)],
a fairly large amount of the plasma heating occurs near the
critical angle $\varphi_c$.  
The parallel component of the electric field in terms of the external
magnetic field is essential for the resonant absorption by electrons
for this case.  
The 1D and 3D PIC results in Fig.~\ref{fig5} make sure that the
angle dependence is unaffected by the spatial dimensions of the simulations. 

In Sec.~\ref{sec4}, the transmittance of the R wave is examined intensively. 
However, as discussed in Sec.~\ref{sec2}, the extraction of only
the L wave may also be possible in the parameter region (II).
A benefit of the L-wave filter is that the
required strength of the magnetic field is relatively smaller, so that it could
be more feasible for practical use. 
On the other hand, a defect of the L wave extraction is a weaker limit in
the laser intensity. 
Because of the relativistic transparency, the cases of $a_0 \gtrsim 1$
are inappropriate for the LCP wave generation by our method. 
Furthermore, it turned out that there is a severe angle dependence in
the transmittance of the L wave.
The ideal transmittance of the L wave is realized only when the
laser injection is strictly parallel to the external magnetic field.
Otherwise, a large fraction of the laser energy goes to the thermal
energy of electrons.

Figure~\ref{fig6} shows the angle dependence of the energy fraction
after the laser-plasma interaction for a typical case in the region (II).
The model parameters calculated here is the plasma density
$\widetilde{n}_e = 1$, the 
magnetic field strength $\widetilde{B}_{\rm ext} = 0.7$, and the
incident laser amplitude $a_0 = 0.1$.
The other parameters are the same as in Fig.~\ref{fig5}.
When the laser injection is parallel to the magnetic field
($\varphi = 0$), both of the transmittance and reflectivity are nearly
half of the injected energy.
In this case, only the L-wave component is transparent, and thus it
works as a converter to LCP waves from an LP laser.
However, if the angle is not zero, the energy absorption by the
electrons cannot be ignored.
If underdense preplasmas form at the ablation side of the target,
the electrons are accelerated effectively by the laser field through 
the $\bm{v \times B}$ force.   
The wave propagation is allowed at any angle for this case, which is
the L wave at $\varphi = 0$ and the X wave at $\varphi = 90$.
The angle dependence of the transmittance and reflectivity exhibits
complicated behavior, as seen in Figs.~\ref{fig6}(a) and \ref{fig6}(b).

\begin{figure}
\includegraphics[scale=0.85,clip]{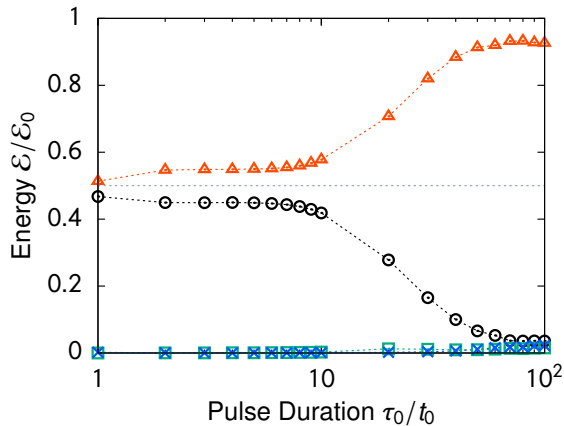}%
\caption{
Dependence of the transmittance and refectivity on the pulse duration
$\tau_0$ of the incident laser obtained by 1D PIC simulations.  
The indicated marks are the same as in Fig.~\ref{fig7}.
The fiducial parameters are used in these runs, but the foil thickness
is assumed to be $L / \lambda_0 = 10$.
The transmittance decreases dramatically if the pulse duration becomes
longer about $\tau_0 / t_0 \gtrsim 10$.
\label{fig8}}
\end{figure}

In this paper, the incident angle of the laser is assumed to be always
normal to the target surface.
Even when the laser has an incident angle to the target, if the
magnetic field is parallel to the laser direction, the transmitted
light should have circular polarization. 
It might be interesting to investigate the dependence of the
transmittance on the laser incident angle together with the $\varphi$
dependence. 
We ignore the effects of preplasmas in this analysis, because the
preplasma has little influence on the propagation of the whistler
wave. 
Throughout this analysis, we assume a relatively shorter pulse
duration for the injected laser, $\widetilde{\tau}_0 =3$.
However, the stimulated Brillouin scattering could prevent
whistler-wave propagation if the pulse duration becomes longer (see
Fig.~\ref{fig8}).
The critical pulse length is determined by the growth time of
the Brillouin instability, and thus it would depend largely on the laser
intensity \cite{sano20}. 
The dependence of the transmittance on the pulse duration would also
be an exciting task to be studied.


\section{Conclusions \label{sec6}}

We have investigated a unique mechanism of CP wave generation by controlling the laser-plasma interaction under a strong magnetic field.
When a thin plasma foil is irradiated by an LP laser under some appropriate conditions, the transmitted and reflected lights become circularly polarized and have nearly the same intensity. 
The whistler-mode of the R wave passes through a thin plasma target when $1 \lesssim \widetilde{B}_{\rm ext} \lesssim \widetilde{n}_e - 1$, while the L-wave component is totally reflected.
To function as a conversion filter, the density of the plasma foil should
be higher than $\widetilde{n}_e > 2$.
The optimal condition requires a strong external magnetic field with a
size of $\widetilde{B}_{\rm ext} \sim \widetilde{n}_e - 1$ along the incident light direction.
The polarity of the CP waves can be switched by the direction of the magnetic field.
The transmittance is unaffected by a small angle difference between the wave injection and the magnetic field less than about 15 degrees.

{\color{black}A significant result of this work is that linear analysis
of this problem is quite adequate for assessing the behavior of this
concept even for the relativistic laser intensity.}
The advantage of this plasma photonic device is a large amplitude of
CP wave, which is valid even for the relativistic amplitude.
Therefore, the high-intensity CP waves generated by this plasma device
could be useful for high-energy-density plasma experiments, such as
laser-driven ion acceleration \cite{esirkepov04,fiuza12} and ion
heating in overdense plasmas \cite{sano19,sano20}.  
It would be possible to propose plasma photonic devices for
various different uses by appropriately selecting the plasma
environment for the density and magnetic field strength.
In the future, it will be meaningful to pursue further device
developments for ultra-intense lasers while utilizing numerical simulations.

\begin{acknowledgments}
We thank Ryosuke Kodama, Kunioki Mima, and Yoshitaka Mori for useful
discussions.   
This research was partially supported by JSPS KAKENHI Grant No. JP19KK0072,
JP20H00140, and JSPS Core-to-Core Program, B. Asia-Africa Science Platforms
No. JPJSCCB20190003.
\end{acknowledgments}



\begin{thebibliography}{28}%
\makeatletter
\providecommand \@ifxundefined [1]{%
 \@ifx{#1\undefined}
}%
\providecommand \@ifnum [1]{%
 \ifnum #1\expandafter \@firstoftwo
 \else \expandafter \@secondoftwo
 \fi
}%
\providecommand \@ifx [1]{%
 \ifx #1\expandafter \@firstoftwo
 \else \expandafter \@secondoftwo
 \fi
}%
\providecommand \natexlab [1]{#1}%
\providecommand \enquote  [1]{``#1''}%
\providecommand \bibnamefont  [1]{#1}%
\providecommand \bibfnamefont [1]{#1}%
\providecommand \citenamefont [1]{#1}%
\providecommand \href@noop [0]{\@secondoftwo}%
\providecommand \href [0]{\begingroup \@sanitize@url \@href}%
\providecommand \@href[1]{\@@startlink{#1}\@@href}%
\providecommand \@@href[1]{\endgroup#1\@@endlink}%
\providecommand \@sanitize@url [0]{\catcode `\\12\catcode `\$12\catcode
  `\&12\catcode `\#12\catcode `\^12\catcode `\_12\catcode `\%12\relax}%
\providecommand \@@startlink[1]{}%
\providecommand \@@endlink[0]{}%
\providecommand \url  [0]{\begingroup\@sanitize@url \@url }%
\providecommand \@url [1]{\endgroup\@href {#1}{\urlprefix }}%
\providecommand \urlprefix  [0]{URL }%
\providecommand \Eprint [0]{\href }%
\providecommand \doibase [0]{http://dx.doi.org/}%
\providecommand \selectlanguage [0]{\@gobble}%
\providecommand \bibinfo  [0]{\@secondoftwo}%
\providecommand \bibfield  [0]{\@secondoftwo}%
\providecommand \translation [1]{[#1]}%
\providecommand \BibitemOpen [0]{}%
\providecommand \bibitemStop [0]{}%
\providecommand \bibitemNoStop [0]{.\EOS\space}%
\providecommand \EOS [0]{\spacefactor3000\relax}%
\providecommand \BibitemShut  [1]{\csname bibitem#1\endcsname}%
\let\auto@bib@innerbib\@empty
\bibitem [{\citenamefont {Kodama}\ \emph {et~al.}(2004)\citenamefont {Kodama},
  \citenamefont {Sentoku}, \citenamefont {Chen}, \citenamefont {Kumar},
  \citenamefont {Hatchett}, \citenamefont {Toyama}, \citenamefont {Cowan},
  \citenamefont {Freeman}, \citenamefont {Fuchs}, \citenamefont {Izawa},
  \citenamefont {Key}, \citenamefont {Kitagawa}, \citenamefont {Kondo},
  \citenamefont {Matsuoka}, \citenamefont {Nakamura}, \citenamefont
  {Nakatsutsumi}, \citenamefont {Norreys}, \citenamefont {Norimatsu},
  \citenamefont {Snavely}, \citenamefont {Stephens}, \citenamefont {Tampo},
  \citenamefont {Tanaka},\ and\ \citenamefont {Yabuuchi}}]{kodama04}%
  \BibitemOpen
  \bibfield  {author} {\bibinfo {author} {\bibfnamefont {R.}~\bibnamefont
  {Kodama}}, \bibinfo {author} {\bibfnamefont {Y.}~\bibnamefont {Sentoku}},
  \bibinfo {author} {\bibfnamefont {Z.~L.}\ \bibnamefont {Chen}}, \bibinfo
  {author} {\bibfnamefont {G.~R.}\ \bibnamefont {Kumar}}, \bibinfo {author}
  {\bibfnamefont {S.~P.}\ \bibnamefont {Hatchett}}, \bibinfo {author}
  {\bibfnamefont {Y.}~\bibnamefont {Toyama}}, \bibinfo {author} {\bibfnamefont
  {T.~E.}\ \bibnamefont {Cowan}}, \bibinfo {author} {\bibfnamefont {R.~R.}\
  \bibnamefont {Freeman}}, \bibinfo {author} {\bibfnamefont {J.}~\bibnamefont
  {Fuchs}}, \bibinfo {author} {\bibfnamefont {Y.}~\bibnamefont {Izawa}},
  \bibinfo {author} {\bibfnamefont {M.~H.}\ \bibnamefont {Key}}, \bibinfo
  {author} {\bibfnamefont {Y.}~\bibnamefont {Kitagawa}}, \bibinfo {author}
  {\bibfnamefont {K.}~\bibnamefont {Kondo}}, \bibinfo {author} {\bibfnamefont
  {T.}~\bibnamefont {Matsuoka}}, \bibinfo {author} {\bibfnamefont
  {H.}~\bibnamefont {Nakamura}}, \bibinfo {author} {\bibfnamefont
  {M.}~\bibnamefont {Nakatsutsumi}}, \bibinfo {author} {\bibfnamefont {P.~A.}\
  \bibnamefont {Norreys}}, \bibinfo {author} {\bibfnamefont {T.}~\bibnamefont
  {Norimatsu}}, \bibinfo {author} {\bibfnamefont {R.~A.}\ \bibnamefont
  {Snavely}}, \bibinfo {author} {\bibfnamefont {R.~B.}\ \bibnamefont
  {Stephens}}, \bibinfo {author} {\bibfnamefont {M.}~\bibnamefont {Tampo}},
  \bibinfo {author} {\bibfnamefont {K.~A.}\ \bibnamefont {Tanaka}}, \ and\
  \bibinfo {author} {\bibfnamefont {T.}~\bibnamefont {Yabuuchi}},\ }\href
  {\doibase 10.1038/nature03133} {\bibfield  {journal} {\bibinfo  {journal}
  {Nature}\ }\textbf {\bibinfo {volume} {432}},\ \bibinfo {pages} {1005}
  (\bibinfo {year} {2004})}\BibitemShut {NoStop}%
\bibitem [{\citenamefont {Kapteyn}\ \emph {et~al.}(1991)\citenamefont
  {Kapteyn}, \citenamefont {Szoke}, \citenamefont {Falcone},\ and\
  \citenamefont {Murnane}}]{kapteyn91}%
  \BibitemOpen
  \bibfield  {author} {\bibinfo {author} {\bibfnamefont {H.~C.}\ \bibnamefont
  {Kapteyn}}, \bibinfo {author} {\bibfnamefont {A.}~\bibnamefont {Szoke}},
  \bibinfo {author} {\bibfnamefont {R.~W.}\ \bibnamefont {Falcone}}, \ and\
  \bibinfo {author} {\bibfnamefont {M.~M.}\ \bibnamefont {Murnane}},\ }\href
  {\doibase 10.1364/OL.16.000490} {\bibfield  {journal} {\bibinfo  {journal}
  {Opt. Lett.}\ }\textbf {\bibinfo {volume} {16}},\ \bibinfo {pages} {490}
  (\bibinfo {year} {1991})}\BibitemShut {NoStop}%
\bibitem [{\citenamefont {Michel}\ \emph {et~al.}(2009)\citenamefont {Michel},
  \citenamefont {Divol}, \citenamefont {Williams}, \citenamefont {Weber},
  \citenamefont {Thomas}, \citenamefont {Callahan}, \citenamefont {Haan},
  \citenamefont {Salmonson}, \citenamefont {Dixit}, \citenamefont {Hinkel},
  \citenamefont {Edwards}, \citenamefont {MacGowan}, \citenamefont {Lindl},
  \citenamefont {Glenzer},\ and\ \citenamefont {Suter}}]{michel09}%
  \BibitemOpen
  \bibfield  {author} {\bibinfo {author} {\bibfnamefont {P.}~\bibnamefont
  {Michel}}, \bibinfo {author} {\bibfnamefont {L.}~\bibnamefont {Divol}},
  \bibinfo {author} {\bibfnamefont {E.~A.}\ \bibnamefont {Williams}}, \bibinfo
  {author} {\bibfnamefont {S.}~\bibnamefont {Weber}}, \bibinfo {author}
  {\bibfnamefont {C.~A.}\ \bibnamefont {Thomas}}, \bibinfo {author}
  {\bibfnamefont {D.~A.}\ \bibnamefont {Callahan}}, \bibinfo {author}
  {\bibfnamefont {S.~W.}\ \bibnamefont {Haan}}, \bibinfo {author}
  {\bibfnamefont {J.~D.}\ \bibnamefont {Salmonson}}, \bibinfo {author}
  {\bibfnamefont {S.}~\bibnamefont {Dixit}}, \bibinfo {author} {\bibfnamefont
  {D.~E.}\ \bibnamefont {Hinkel}}, \bibinfo {author} {\bibfnamefont {M.~J.}\
  \bibnamefont {Edwards}}, \bibinfo {author} {\bibfnamefont {B.~J.}\
  \bibnamefont {MacGowan}}, \bibinfo {author} {\bibfnamefont {J.~D.}\
  \bibnamefont {Lindl}}, \bibinfo {author} {\bibfnamefont {S.~H.}\ \bibnamefont
  {Glenzer}}, \ and\ \bibinfo {author} {\bibfnamefont {L.~J.}\ \bibnamefont
  {Suter}},\ }\href {\doibase 10.1103/PhysRevLett.102.025004} {\bibfield
  {journal} {\bibinfo  {journal} {Phys. Rev. Lett.}\ }\textbf {\bibinfo
  {volume} {102}},\ \bibinfo {pages} {025004} (\bibinfo {year}
  {2009})}\BibitemShut {NoStop}%
\bibitem [{\citenamefont {Moody}\ \emph {et~al.}(2012)\citenamefont {Moody},
  \citenamefont {Michel}, \citenamefont {Divol}, \citenamefont {Berger},
  \citenamefont {Bond}, \citenamefont {Bradley}, \citenamefont {Callahan},
  \citenamefont {Dewald}, \citenamefont {Dixit}, \citenamefont {Edwards},
  \citenamefont {Glenn}, \citenamefont {Hamza}, \citenamefont {Haynam},
  \citenamefont {Hinkel}, \citenamefont {Izumi}, \citenamefont {Jones},
  \citenamefont {Kilkenny}, \citenamefont {Kirkwood}, \citenamefont {Kline},
  \citenamefont {Kruer}, \citenamefont {Kyrala}, \citenamefont {Landen},
  \citenamefont {LePape}, \citenamefont {Lindl}, \citenamefont {MacGowan},
  \citenamefont {Meezan}, \citenamefont {Nikroo}, \citenamefont {Rosen},
  \citenamefont {Schneider}, \citenamefont {Strozzi}, \citenamefont {Suter},
  \citenamefont {Thomas}, \citenamefont {Town}, \citenamefont {Widmann},
  \citenamefont {Williams}, \citenamefont {Atherton}, \citenamefont {Glenzer},\
  and\ \citenamefont {Moses}}]{moody12}%
  \BibitemOpen
  \bibfield  {author} {\bibinfo {author} {\bibfnamefont {J.~D.}\ \bibnamefont
  {Moody}}, \bibinfo {author} {\bibfnamefont {P.}~\bibnamefont {Michel}},
  \bibinfo {author} {\bibfnamefont {L.}~\bibnamefont {Divol}}, \bibinfo
  {author} {\bibfnamefont {R.~L.}\ \bibnamefont {Berger}}, \bibinfo {author}
  {\bibfnamefont {E.}~\bibnamefont {Bond}}, \bibinfo {author} {\bibfnamefont
  {D.~K.}\ \bibnamefont {Bradley}}, \bibinfo {author} {\bibfnamefont {D.~A.}\
  \bibnamefont {Callahan}}, \bibinfo {author} {\bibfnamefont {E.~L.}\
  \bibnamefont {Dewald}}, \bibinfo {author} {\bibfnamefont {S.}~\bibnamefont
  {Dixit}}, \bibinfo {author} {\bibfnamefont {M.~J.}\ \bibnamefont {Edwards}},
  \bibinfo {author} {\bibfnamefont {S.}~\bibnamefont {Glenn}}, \bibinfo
  {author} {\bibfnamefont {A.}~\bibnamefont {Hamza}}, \bibinfo {author}
  {\bibfnamefont {C.}~\bibnamefont {Haynam}}, \bibinfo {author} {\bibfnamefont
  {D.~E.}\ \bibnamefont {Hinkel}}, \bibinfo {author} {\bibfnamefont
  {N.}~\bibnamefont {Izumi}}, \bibinfo {author} {\bibfnamefont
  {O.}~\bibnamefont {Jones}}, \bibinfo {author} {\bibfnamefont {J.~D.}\
  \bibnamefont {Kilkenny}}, \bibinfo {author} {\bibfnamefont {R.~K.}\
  \bibnamefont {Kirkwood}}, \bibinfo {author} {\bibfnamefont {J.~L.}\
  \bibnamefont {Kline}}, \bibinfo {author} {\bibfnamefont {W.~L.}\ \bibnamefont
  {Kruer}}, \bibinfo {author} {\bibfnamefont {G.~A.}\ \bibnamefont {Kyrala}},
  \bibinfo {author} {\bibfnamefont {O.~L.}\ \bibnamefont {Landen}}, \bibinfo
  {author} {\bibfnamefont {S.}~\bibnamefont {LePape}}, \bibinfo {author}
  {\bibfnamefont {J.~D.}\ \bibnamefont {Lindl}}, \bibinfo {author}
  {\bibfnamefont {B.~J.}\ \bibnamefont {MacGowan}}, \bibinfo {author}
  {\bibfnamefont {N.~B.}\ \bibnamefont {Meezan}}, \bibinfo {author}
  {\bibfnamefont {A.}~\bibnamefont {Nikroo}}, \bibinfo {author} {\bibfnamefont
  {M.~D.}\ \bibnamefont {Rosen}}, \bibinfo {author} {\bibfnamefont {M.~B.}\
  \bibnamefont {Schneider}}, \bibinfo {author} {\bibfnamefont {D.~J.}\
  \bibnamefont {Strozzi}}, \bibinfo {author} {\bibfnamefont {L.~J.}\
  \bibnamefont {Suter}}, \bibinfo {author} {\bibfnamefont {C.~A.}\ \bibnamefont
  {Thomas}}, \bibinfo {author} {\bibfnamefont {R.~P.~J.}\ \bibnamefont {Town}},
  \bibinfo {author} {\bibfnamefont {K.}~\bibnamefont {Widmann}}, \bibinfo
  {author} {\bibfnamefont {E.~A.}\ \bibnamefont {Williams}}, \bibinfo {author}
  {\bibfnamefont {L.~J.}\ \bibnamefont {Atherton}}, \bibinfo {author}
  {\bibfnamefont {S.~H.}\ \bibnamefont {Glenzer}}, \ and\ \bibinfo {author}
  {\bibfnamefont {E.~I.}\ \bibnamefont {Moses}},\ }\href {\doibase
  10.1038/nphys2239} {\bibfield  {journal} {\bibinfo  {journal} {Nat. Phys.}\
  }\textbf {\bibinfo {volume} {8}},\ \bibinfo {pages} {344} (\bibinfo {year}
  {2012})}\BibitemShut {NoStop}%
\bibitem [{\citenamefont {Malkin}\ \emph {et~al.}(1999)\citenamefont {Malkin},
  \citenamefont {Shvets},\ and\ \citenamefont {Fisch}}]{malkin99}%
  \BibitemOpen
  \bibfield  {author} {\bibinfo {author} {\bibfnamefont {V.~M.}\ \bibnamefont
  {Malkin}}, \bibinfo {author} {\bibfnamefont {G.}~\bibnamefont {Shvets}}, \
  and\ \bibinfo {author} {\bibfnamefont {N.~J.}\ \bibnamefont {Fisch}},\ }\href
  {\doibase 10.1103/PhysRevLett.82.4448} {\bibfield  {journal} {\bibinfo
  {journal} {Phys.l Rev. Lett.}\ }\textbf {\bibinfo {volume} {82}},\ \bibinfo
  {pages} {4448} (\bibinfo {year} {1999})}\BibitemShut {NoStop}%
\bibitem [{\citenamefont {Shi}\ \emph {et~al.}(2017)\citenamefont {Shi},
  \citenamefont {Qin},\ and\ \citenamefont {Fisch}}]{shi17}%
  \BibitemOpen
  \bibfield  {author} {\bibinfo {author} {\bibfnamefont {Y.}~\bibnamefont
  {Shi}}, \bibinfo {author} {\bibfnamefont {H.}~\bibnamefont {Qin}}, \ and\
  \bibinfo {author} {\bibfnamefont {N.~J.}\ \bibnamefont {Fisch}},\ }\href
  {\doibase 10.1103/PhysRevE.95.023211} {\bibfield  {journal} {\bibinfo
  {journal} {Phys. Rev. E}\ }\textbf {\bibinfo {volume} {95}},\ \bibinfo
  {pages} {023211} (\bibinfo {year} {2017})}\BibitemShut {NoStop}%
\bibitem [{\citenamefont {Carman}\ \emph {et~al.}(1981)\citenamefont {Carman},
  \citenamefont {Forslund},\ and\ \citenamefont {Kindel}}]{carman81}%
  \BibitemOpen
  \bibfield  {author} {\bibinfo {author} {\bibfnamefont {R.~L.}\ \bibnamefont
  {Carman}}, \bibinfo {author} {\bibfnamefont {D.~W.}\ \bibnamefont
  {Forslund}}, \ and\ \bibinfo {author} {\bibfnamefont {J.~M.}\ \bibnamefont
  {Kindel}},\ }\href {\doibase 10.1103/PhysRevLett.46.29} {\bibfield  {journal}
  {\bibinfo  {journal} {Phys. Rev. Lett.}\ }\textbf {\bibinfo {volume} {46}},\
  \bibinfo {pages} {29} (\bibinfo {year} {1981})}\BibitemShut {NoStop}%
\bibitem [{\citenamefont {{Esirkepov}}\ \emph {et~al.}(2004)\citenamefont
  {{Esirkepov}}, \citenamefont {{Borghesi}}, \citenamefont {{Bulanov}},
  \citenamefont {{Mourou}},\ and\ \citenamefont {{Tajima}}}]{esirkepov04}%
  \BibitemOpen
  \bibfield  {author} {\bibinfo {author} {\bibfnamefont {T.}~\bibnamefont
  {{Esirkepov}}}, \bibinfo {author} {\bibfnamefont {M.}~\bibnamefont
  {{Borghesi}}}, \bibinfo {author} {\bibfnamefont {S.~V.}\ \bibnamefont
  {{Bulanov}}}, \bibinfo {author} {\bibfnamefont {G.}~\bibnamefont {{Mourou}}},
  \ and\ \bibinfo {author} {\bibfnamefont {T.}~\bibnamefont {{Tajima}}},\
  }\href {\doibase 10.1103/PhysRevLett.92.175003} {\bibfield  {journal}
  {\bibinfo  {journal} {Phys. Rev. Lett.}\ }\textbf {\bibinfo {volume} {92}},\
  \bibinfo {eid} {175003} (\bibinfo {year} {2004})}\BibitemShut {NoStop}%
\bibitem [{\citenamefont {Fiuza}\ \emph {et~al.}(2012)\citenamefont {Fiuza},
  \citenamefont {Stockem}, \citenamefont {Boella}, \citenamefont {Fonseca},
  \citenamefont {Silva}, \citenamefont {Haberberger}, \citenamefont
  {Tochitsky}, \citenamefont {Gong}, \citenamefont {Mori},\ and\ \citenamefont
  {Joshi}}]{fiuza12}%
  \BibitemOpen
  \bibfield  {author} {\bibinfo {author} {\bibfnamefont {F.}~\bibnamefont
  {Fiuza}}, \bibinfo {author} {\bibfnamefont {A.}~\bibnamefont {Stockem}},
  \bibinfo {author} {\bibfnamefont {E.}~\bibnamefont {Boella}}, \bibinfo
  {author} {\bibfnamefont {R.~A.}\ \bibnamefont {Fonseca}}, \bibinfo {author}
  {\bibfnamefont {L.~O.}\ \bibnamefont {Silva}}, \bibinfo {author}
  {\bibfnamefont {D.}~\bibnamefont {Haberberger}}, \bibinfo {author}
  {\bibfnamefont {S.}~\bibnamefont {Tochitsky}}, \bibinfo {author}
  {\bibfnamefont {C.}~\bibnamefont {Gong}}, \bibinfo {author} {\bibfnamefont
  {W.~B.}\ \bibnamefont {Mori}}, \ and\ \bibinfo {author} {\bibfnamefont
  {C.}~\bibnamefont {Joshi}},\ }\href {\doibase 10.1103/PhysRevLett.109.215001}
  {\bibfield  {journal} {\bibinfo  {journal} {Phys. Rev. Lett.}\ }\textbf
  {\bibinfo {volume} {109}},\ \bibinfo {pages} {215001} (\bibinfo {year}
  {2012})}\BibitemShut {NoStop}%
\bibitem [{\citenamefont {Sano}\ \emph {et~al.}(2019)\citenamefont {Sano},
  \citenamefont {Hata}, \citenamefont {Kawahito}, \citenamefont {Mima},\ and\
  \citenamefont {Sentoku}}]{sano19}%
  \BibitemOpen
  \bibfield  {author} {\bibinfo {author} {\bibfnamefont {T.}~\bibnamefont
  {Sano}}, \bibinfo {author} {\bibfnamefont {M.}~\bibnamefont {Hata}}, \bibinfo
  {author} {\bibfnamefont {D.}~\bibnamefont {Kawahito}}, \bibinfo {author}
  {\bibfnamefont {K.}~\bibnamefont {Mima}}, \ and\ \bibinfo {author}
  {\bibfnamefont {Y.}~\bibnamefont {Sentoku}},\ }\href {\doibase
  10.1103/PhysRevE.100.053205} {\bibfield  {journal} {\bibinfo  {journal}
  {Phys. Rev. E}\ }\textbf {\bibinfo {volume} {100}},\ \bibinfo {pages}
  {053205} (\bibinfo {year} {2019})}\BibitemShut {NoStop}%
\bibitem [{\citenamefont {Sano}\ \emph {et~al.}(2020)\citenamefont {Sano},
  \citenamefont {Fujioka}, \citenamefont {Mori}, \citenamefont {Mima},\ and\
  \citenamefont {Sentoku}}]{sano20}%
  \BibitemOpen
  \bibfield  {author} {\bibinfo {author} {\bibfnamefont {T.}~\bibnamefont
  {Sano}}, \bibinfo {author} {\bibfnamefont {S.}~\bibnamefont {Fujioka}},
  \bibinfo {author} {\bibfnamefont {Y.}~\bibnamefont {Mori}}, \bibinfo {author}
  {\bibfnamefont {K.}~\bibnamefont {Mima}}, \ and\ \bibinfo {author}
  {\bibfnamefont {Y.}~\bibnamefont {Sentoku}},\ }\href {\doibase
  10.1103/PhysRevE.101.013206} {\bibfield  {journal} {\bibinfo  {journal}
  {Phys. Rev. E}\ }\textbf {\bibinfo {volume} {101}},\ \bibinfo {pages}
  {013206} (\bibinfo {year} {2020})}\BibitemShut {NoStop}%
\bibitem [{\citenamefont {Michel}\ \emph {et~al.}(2014)\citenamefont {Michel},
  \citenamefont {Divol}, \citenamefont {Turnbull},\ and\ \citenamefont
  {Moody}}]{michel14}%
  \BibitemOpen
  \bibfield  {author} {\bibinfo {author} {\bibfnamefont {P.}~\bibnamefont
  {Michel}}, \bibinfo {author} {\bibfnamefont {L.}~\bibnamefont {Divol}},
  \bibinfo {author} {\bibfnamefont {D.}~\bibnamefont {Turnbull}}, \ and\
  \bibinfo {author} {\bibfnamefont {J.~D.}\ \bibnamefont {Moody}},\ }\href
  {\doibase 10.1103/PhysRevLett.113.205001} {\bibfield  {journal} {\bibinfo
  {journal} {Phys. Rev. Lett.}\ }\textbf {\bibinfo {volume} {113}},\ \bibinfo
  {pages} {205001} (\bibinfo {year} {2014})}\BibitemShut {NoStop}%
\bibitem [{\citenamefont {Michel}\ \emph {et~al.}(2020)\citenamefont {Michel},
  \citenamefont {Kur}, \citenamefont {Lazarow}, \citenamefont {Chapman},
  \citenamefont {Divol},\ and\ \citenamefont {Wurtele}}]{michel20}%
  \BibitemOpen
  \bibfield  {author} {\bibinfo {author} {\bibfnamefont {P.}~\bibnamefont
  {Michel}}, \bibinfo {author} {\bibfnamefont {E.}~\bibnamefont {Kur}},
  \bibinfo {author} {\bibfnamefont {M.}~\bibnamefont {Lazarow}}, \bibinfo
  {author} {\bibfnamefont {T.}~\bibnamefont {Chapman}}, \bibinfo {author}
  {\bibfnamefont {L.}~\bibnamefont {Divol}}, \ and\ \bibinfo {author}
  {\bibfnamefont {J.~S.}\ \bibnamefont {Wurtele}},\ }\href {\doibase
  10.1103/PhysRevX.10.021039} {\bibfield  {journal} {\bibinfo  {journal} {Phys.
  Rev. X}\ }\textbf {\bibinfo {volume} {10}},\ \bibinfo {pages} {021039}
  (\bibinfo {year} {2020})}\BibitemShut {NoStop}%
\bibitem [{\citenamefont {Lehmann}\ and\ \citenamefont
  {Spatschek}(2016)}]{lehmann16}%
  \BibitemOpen
  \bibfield  {author} {\bibinfo {author} {\bibfnamefont {G.}~\bibnamefont
  {Lehmann}}\ and\ \bibinfo {author} {\bibfnamefont {K.~H.}\ \bibnamefont
  {Spatschek}},\ }\href {\doibase 10.1103/PhysRevLett.116.225002} {\bibfield
  {journal} {\bibinfo  {journal} {Phys. Rev. Lett.}\ }\textbf {\bibinfo
  {volume} {116}},\ \bibinfo {pages} {225002} (\bibinfo {year}
  {2016})}\BibitemShut {NoStop}%
\bibitem [{\citenamefont {Lehmann}\ and\ \citenamefont
  {Spatschek}(2018)}]{lehmann18}%
  \BibitemOpen
  \bibfield  {author} {\bibinfo {author} {\bibfnamefont {G.}~\bibnamefont
  {Lehmann}}\ and\ \bibinfo {author} {\bibfnamefont {K.~H.}\ \bibnamefont
  {Spatschek}},\ }\href {\doibase 10.1103/PhysRevE.97.063201} {\bibfield
  {journal} {\bibinfo  {journal} {Phys. Rev. E}\ }\textbf {\bibinfo {volume}
  {97}},\ \bibinfo {pages} {063201} (\bibinfo {year} {2018})}\BibitemShut
  {NoStop}%
\bibitem [{\citenamefont {Stark}\ \emph {et~al.}(2015)\citenamefont {Stark},
  \citenamefont {Bhattacharjee}, \citenamefont {Arefiev}, \citenamefont
  {Toncian}, \citenamefont {Hazeltine},\ and\ \citenamefont
  {Mahajan}}]{stark15}%
  \BibitemOpen
  \bibfield  {author} {\bibinfo {author} {\bibfnamefont {D.~J.}\ \bibnamefont
  {Stark}}, \bibinfo {author} {\bibfnamefont {C.}~\bibnamefont
  {Bhattacharjee}}, \bibinfo {author} {\bibfnamefont {A.~V.}\ \bibnamefont
  {Arefiev}}, \bibinfo {author} {\bibfnamefont {T.}~\bibnamefont {Toncian}},
  \bibinfo {author} {\bibfnamefont {R.~D.}\ \bibnamefont {Hazeltine}}, \ and\
  \bibinfo {author} {\bibfnamefont {S.~M.}\ \bibnamefont {Mahajan}},\ }\href
  {\doibase 10.1103/PhysRevLett.115.025002} {\bibfield  {journal} {\bibinfo
  {journal} {Phys. Rev. Lett.}\ }\textbf {\bibinfo {volume} {115}},\ \bibinfo
  {pages} {025002} (\bibinfo {year} {2015})},\ \Eprint
  {http://arxiv.org/abs/1412.1865} {1412.1865} \BibitemShut {NoStop}%
\bibitem [{\citenamefont {{Fujioka}}\ \emph {et~al.}(2013)\citenamefont
  {{Fujioka}}, \citenamefont {{Zhang}}, \citenamefont {{Ishihara}},
  \citenamefont {{Shigemori}}, \citenamefont {{Hironaka}}, \citenamefont
  {{Johzaki}}, \citenamefont {{Sunahara}}, \citenamefont {{Yamamoto}},
  \citenamefont {{Nakashima}}, \citenamefont {{Watanabe}}, \citenamefont
  {{Shiraga}}, \citenamefont {{Nishimura}},\ and\ \citenamefont
  {{Azechi}}}]{fujioka13}%
  \BibitemOpen
  \bibfield  {author} {\bibinfo {author} {\bibfnamefont {S.}~\bibnamefont
  {{Fujioka}}}, \bibinfo {author} {\bibfnamefont {Z.}~\bibnamefont {{Zhang}}},
  \bibinfo {author} {\bibfnamefont {K.}~\bibnamefont {{Ishihara}}}, \bibinfo
  {author} {\bibfnamefont {K.}~\bibnamefont {{Shigemori}}}, \bibinfo {author}
  {\bibfnamefont {Y.}~\bibnamefont {{Hironaka}}}, \bibinfo {author}
  {\bibfnamefont {T.}~\bibnamefont {{Johzaki}}}, \bibinfo {author}
  {\bibfnamefont {A.}~\bibnamefont {{Sunahara}}}, \bibinfo {author}
  {\bibfnamefont {N.}~\bibnamefont {{Yamamoto}}}, \bibinfo {author}
  {\bibfnamefont {H.}~\bibnamefont {{Nakashima}}}, \bibinfo {author}
  {\bibfnamefont {T.}~\bibnamefont {{Watanabe}}}, \bibinfo {author}
  {\bibfnamefont {H.}~\bibnamefont {{Shiraga}}}, \bibinfo {author}
  {\bibfnamefont {H.}~\bibnamefont {{Nishimura}}}, \ and\ \bibinfo {author}
  {\bibfnamefont {H.}~\bibnamefont {{Azechi}}},\ }\href@noop {} {\bibfield
  {journal} {\bibinfo  {journal} {Sci. Rep.}\ }\textbf {\bibinfo {volume}
  {3}},\ \bibinfo {pages} {1170} (\bibinfo {year} {2013})}\BibitemShut
  {NoStop}%
\bibitem [{\citenamefont {{Santos}}\ \emph {et~al.}(2018)\citenamefont
  {{Santos}}, \citenamefont {{Bailly-Grandvaux}}, \citenamefont {{Ehret}},
  \citenamefont {{Arefiev}}, \citenamefont {{Batani}}, \citenamefont {{Beg}},
  \citenamefont {{Calisti}}, \citenamefont {{Ferri}}, \citenamefont
  {{Florido}}, \citenamefont {{Forestier-Colleoni}}, \citenamefont {{Fujioka}},
  \citenamefont {{Gigosos}}, \citenamefont {{Giuffrida}}, \citenamefont
  {{Gremillet}}, \citenamefont {{Honrubia}}, \citenamefont {{Kojima}},
  \citenamefont {{Korneev}}, \citenamefont {{Law}}, \citenamefont
  {{Marqu{\`e}s}}, \citenamefont {{Morace}}, \citenamefont {{Moss{\'e}}},
  \citenamefont {{Peyrusse}}, \citenamefont {{Rose}}, \citenamefont {{Roth}},
  \citenamefont {{Sakata}}, \citenamefont {{Schaumann}}, \citenamefont
  {{Suzuki-Vidal}}, \citenamefont {{Tikhonchuk}}, \citenamefont {{Toncian}},
  \citenamefont {{Woolsey}},\ and\ \citenamefont {{Zhang}}}]{santos18}%
  \BibitemOpen
  \bibfield  {author} {\bibinfo {author} {\bibfnamefont {J.~J.}\ \bibnamefont
  {{Santos}}}, \bibinfo {author} {\bibfnamefont {M.}~\bibnamefont
  {{Bailly-Grandvaux}}}, \bibinfo {author} {\bibfnamefont {M.}~\bibnamefont
  {{Ehret}}}, \bibinfo {author} {\bibfnamefont {A.~V.}\ \bibnamefont
  {{Arefiev}}}, \bibinfo {author} {\bibfnamefont {D.}~\bibnamefont {{Batani}}},
  \bibinfo {author} {\bibfnamefont {F.~N.}\ \bibnamefont {{Beg}}}, \bibinfo
  {author} {\bibfnamefont {A.}~\bibnamefont {{Calisti}}}, \bibinfo {author}
  {\bibfnamefont {S.}~\bibnamefont {{Ferri}}}, \bibinfo {author} {\bibfnamefont
  {R.}~\bibnamefont {{Florido}}}, \bibinfo {author} {\bibfnamefont
  {P.}~\bibnamefont {{Forestier-Colleoni}}}, \bibinfo {author} {\bibfnamefont
  {S.}~\bibnamefont {{Fujioka}}}, \bibinfo {author} {\bibfnamefont {M.~A.}\
  \bibnamefont {{Gigosos}}}, \bibinfo {author} {\bibfnamefont {L.}~\bibnamefont
  {{Giuffrida}}}, \bibinfo {author} {\bibfnamefont {L.}~\bibnamefont
  {{Gremillet}}}, \bibinfo {author} {\bibfnamefont {J.~J.}\ \bibnamefont
  {{Honrubia}}}, \bibinfo {author} {\bibfnamefont {S.}~\bibnamefont
  {{Kojima}}}, \bibinfo {author} {\bibfnamefont {P.}~\bibnamefont {{Korneev}}},
  \bibinfo {author} {\bibfnamefont {K.~F.~F.}\ \bibnamefont {{Law}}}, \bibinfo
  {author} {\bibfnamefont {J.-R.}\ \bibnamefont {{Marqu{\`e}s}}}, \bibinfo
  {author} {\bibfnamefont {A.}~\bibnamefont {{Morace}}}, \bibinfo {author}
  {\bibfnamefont {C.}~\bibnamefont {{Moss{\'e}}}}, \bibinfo {author}
  {\bibfnamefont {O.}~\bibnamefont {{Peyrusse}}}, \bibinfo {author}
  {\bibfnamefont {S.}~\bibnamefont {{Rose}}}, \bibinfo {author} {\bibfnamefont
  {M.}~\bibnamefont {{Roth}}}, \bibinfo {author} {\bibfnamefont
  {S.}~\bibnamefont {{Sakata}}}, \bibinfo {author} {\bibfnamefont
  {G.}~\bibnamefont {{Schaumann}}}, \bibinfo {author} {\bibfnamefont
  {F.}~\bibnamefont {{Suzuki-Vidal}}}, \bibinfo {author} {\bibfnamefont
  {V.~T.}\ \bibnamefont {{Tikhonchuk}}}, \bibinfo {author} {\bibfnamefont
  {T.}~\bibnamefont {{Toncian}}}, \bibinfo {author} {\bibfnamefont
  {N.}~\bibnamefont {{Woolsey}}}, \ and\ \bibinfo {author} {\bibfnamefont
  {Z.}~\bibnamefont {{Zhang}}},\ }\href {\doibase 10.1063/1.5018735} {\bibfield
   {journal} {\bibinfo  {journal} {Phys. Plasmas}\ }\textbf {\bibinfo {volume}
  {25}},\ \bibinfo {eid} {056705} (\bibinfo {year} {2018})}\BibitemShut
  {NoStop}%
\bibitem [{\citenamefont {{Budden}}(1961)}]{budden61}%
  \BibitemOpen
  \bibfield  {author} {\bibinfo {author} {\bibfnamefont {K.~G.}\ \bibnamefont
  {{Budden}}},\ }\href@noop {} {\emph {\bibinfo {title} {{Radio waves in the
  ionosphere}}}}\ (\bibinfo  {publisher} {Cambridge University Press},\
  \bibinfo {address} {New York},\ \bibinfo {year} {1961})\BibitemShut {NoStop}%
\bibitem [{\citenamefont {{Yang}}\ \emph {et~al.}(2015)\citenamefont {{Yang}},
  \citenamefont {{Yu}}, \citenamefont {{Xu}}, \citenamefont {{Yu}},
  \citenamefont {{Ge}}, \citenamefont {{Xu}}, \citenamefont {{Zhuo}},
  \citenamefont {{Ma}}, \citenamefont {{Shao}},\ and\ \citenamefont
  {{Borghesi}}}]{yang15}%
  \BibitemOpen
  \bibfield  {author} {\bibinfo {author} {\bibfnamefont {X.~H.}\ \bibnamefont
  {{Yang}}}, \bibinfo {author} {\bibfnamefont {W.}~\bibnamefont {{Yu}}},
  \bibinfo {author} {\bibfnamefont {H.}~\bibnamefont {{Xu}}}, \bibinfo {author}
  {\bibfnamefont {M.~Y.}\ \bibnamefont {{Yu}}}, \bibinfo {author}
  {\bibfnamefont {Z.~Y.}\ \bibnamefont {{Ge}}}, \bibinfo {author}
  {\bibfnamefont {B.~B.}\ \bibnamefont {{Xu}}}, \bibinfo {author}
  {\bibfnamefont {H.~B.}\ \bibnamefont {{Zhuo}}}, \bibinfo {author}
  {\bibfnamefont {Y.~Y.}\ \bibnamefont {{Ma}}}, \bibinfo {author}
  {\bibfnamefont {F.~Q.}\ \bibnamefont {{Shao}}}, \ and\ \bibinfo {author}
  {\bibfnamefont {M.}~\bibnamefont {{Borghesi}}},\ }\href {\doibase
  10.1063/1.4922228} {\bibfield  {journal} {\bibinfo  {journal} {Appl. Phys.
  Lett.}\ }\textbf {\bibinfo {volume} {106}},\ \bibinfo {eid} {224103}
  (\bibinfo {year} {2015})}\BibitemShut {NoStop}%
\bibitem [{\citenamefont {Luan}\ \emph {et~al.}(2016)\citenamefont {Luan},
  \citenamefont {Yu}, \citenamefont {Li}, \citenamefont {Wu}, \citenamefont
  {Sheng}, \citenamefont {Yu},\ and\ \citenamefont {Zhang}}]{luan16}%
  \BibitemOpen
  \bibfield  {author} {\bibinfo {author} {\bibfnamefont {S.~X.}\ \bibnamefont
  {Luan}}, \bibinfo {author} {\bibfnamefont {W.}~\bibnamefont {Yu}}, \bibinfo
  {author} {\bibfnamefont {F.~Y.}\ \bibnamefont {Li}}, \bibinfo {author}
  {\bibfnamefont {D.}~\bibnamefont {Wu}}, \bibinfo {author} {\bibfnamefont
  {Z.~M.}\ \bibnamefont {Sheng}}, \bibinfo {author} {\bibfnamefont {M.~Y.}\
  \bibnamefont {Yu}}, \ and\ \bibinfo {author} {\bibfnamefont {J.}~\bibnamefont
  {Zhang}},\ }\href {\doibase 10.1103/PhysRevE.94.053207} {\bibfield  {journal}
  {\bibinfo  {journal} {Phys. Rev. E}\ }\textbf {\bibinfo {volume} {94}},\
  \bibinfo {pages} {053207} (\bibinfo {year} {2016})}\BibitemShut {NoStop}%
\bibitem [{\citenamefont {Ma}\ \emph {et~al.}(2016)\citenamefont {Ma},
  \citenamefont {Yu}, \citenamefont {Yu}, \citenamefont {Luan},\ and\
  \citenamefont {Wu}}]{ma16}%
  \BibitemOpen
  \bibfield  {author} {\bibinfo {author} {\bibfnamefont {G.}~\bibnamefont
  {Ma}}, \bibinfo {author} {\bibfnamefont {W.}~\bibnamefont {Yu}}, \bibinfo
  {author} {\bibfnamefont {M.~Y.}\ \bibnamefont {Yu}}, \bibinfo {author}
  {\bibfnamefont {S.}~\bibnamefont {Luan}}, \ and\ \bibinfo {author}
  {\bibfnamefont {D.}~\bibnamefont {Wu}},\ }\href {\doibase
  10.1103/PhysRevE.93.053209} {\bibfield  {journal} {\bibinfo  {journal} {Phys.
  Rev. E}\ }\textbf {\bibinfo {volume} {93}},\ \bibinfo {pages} {053209}
  (\bibinfo {year} {2016})}\BibitemShut {NoStop}%
\bibitem [{\citenamefont {Weng}\ \emph {et~al.}(2017)\citenamefont {Weng},
  \citenamefont {Zhao}, \citenamefont {Sheng}, \citenamefont {Yu},
  \citenamefont {Luan}, \citenamefont {Chen}, \citenamefont {Yu}, \citenamefont
  {Murakami}, \citenamefont {Mori},\ and\ \citenamefont {Zhang}}]{weng17}%
  \BibitemOpen
  \bibfield  {author} {\bibinfo {author} {\bibfnamefont {S.}~\bibnamefont
  {Weng}}, \bibinfo {author} {\bibfnamefont {Q.}~\bibnamefont {Zhao}}, \bibinfo
  {author} {\bibfnamefont {Z.}~\bibnamefont {Sheng}}, \bibinfo {author}
  {\bibfnamefont {W.}~\bibnamefont {Yu}}, \bibinfo {author} {\bibfnamefont
  {S.}~\bibnamefont {Luan}}, \bibinfo {author} {\bibfnamefont {M.}~\bibnamefont
  {Chen}}, \bibinfo {author} {\bibfnamefont {L.}~\bibnamefont {Yu}}, \bibinfo
  {author} {\bibfnamefont {M.}~\bibnamefont {Murakami}}, \bibinfo {author}
  {\bibfnamefont {W.~B.}\ \bibnamefont {Mori}}, \ and\ \bibinfo {author}
  {\bibfnamefont {J.}~\bibnamefont {Zhang}},\ }\href {\doibase
  10.1364/OPTICA.4.001086} {\bibfield  {journal} {\bibinfo  {journal} {Optica}\
  }\textbf {\bibinfo {volume} {4}},\ \bibinfo {pages} {1086} (\bibinfo {year}
  {2017})}\BibitemShut {NoStop}%
\bibitem [{\citenamefont {{Stix}}(1992)}]{stix92}%
  \BibitemOpen
  \bibfield  {author} {\bibinfo {author} {\bibfnamefont {T.~H.}\ \bibnamefont
  {{Stix}}},\ }\href@noop {} {\emph {\bibinfo {title} {{Waves in plasmas}}}}\
  (\bibinfo  {publisher} {Springer-Verlag},\ \bibinfo {address} {New York},\
  \bibinfo {year} {1992})\BibitemShut {NoStop}%
\bibitem [{\citenamefont {{Sentoku}}\ and\ \citenamefont
  {{Kemp}}(2008)}]{sentoku08}%
  \BibitemOpen
  \bibfield  {author} {\bibinfo {author} {\bibfnamefont {Y.}~\bibnamefont
  {{Sentoku}}}\ and\ \bibinfo {author} {\bibfnamefont {A.~J.}\ \bibnamefont
  {{Kemp}}},\ }\href {\doibase 10.1016/j.jcp.2008.03.043} {\bibfield  {journal}
  {\bibinfo  {journal} {J. Comp. Phys.}\ }\textbf {\bibinfo {volume} {227}},\
  \bibinfo {pages} {6846} (\bibinfo {year} {2008})}\BibitemShut {NoStop}%
\bibitem [{\citenamefont {Sano}\ \emph {et~al.}(2017)\citenamefont {Sano},
  \citenamefont {Tanaka}, \citenamefont {Iwata}, \citenamefont {Hata},
  \citenamefont {Mima}, \citenamefont {Murakami},\ and\ \citenamefont
  {Sentoku}}]{sano17}%
  \BibitemOpen
  \bibfield  {author} {\bibinfo {author} {\bibfnamefont {T.}~\bibnamefont
  {Sano}}, \bibinfo {author} {\bibfnamefont {Y.}~\bibnamefont {Tanaka}},
  \bibinfo {author} {\bibfnamefont {N.}~\bibnamefont {Iwata}}, \bibinfo
  {author} {\bibfnamefont {M.}~\bibnamefont {Hata}}, \bibinfo {author}
  {\bibfnamefont {K.}~\bibnamefont {Mima}}, \bibinfo {author} {\bibfnamefont
  {M.}~\bibnamefont {Murakami}}, \ and\ \bibinfo {author} {\bibfnamefont
  {Y.}~\bibnamefont {Sentoku}},\ }\href {\doibase 10.1103/PhysRevE.96.043209}
  {\bibfield  {journal} {\bibinfo  {journal} {Phys. Rev. E}\ }\textbf {\bibinfo
  {volume} {96}},\ \bibinfo {pages} {043209} (\bibinfo {year}
  {2017})}\BibitemShut {NoStop}%
\bibitem [{\citenamefont {Kaw}\ and\ \citenamefont {Dawson}(1970)}]{kaw70}%
  \BibitemOpen
  \bibfield  {author} {\bibinfo {author} {\bibfnamefont {P.}~\bibnamefont
  {Kaw}}\ and\ \bibinfo {author} {\bibfnamefont {J.}~\bibnamefont {Dawson}},\
  }\href {\doibase 10.1063/1.1692942} {\bibfield  {journal} {\bibinfo
  {journal} {Phys. Fluids}\ }\textbf {\bibinfo {volume} {13}},\ \bibinfo
  {pages} {472} (\bibinfo {year} {1970})}\BibitemShut {NoStop}%
\bibitem [{\citenamefont {Sakagami}\ and\ \citenamefont
  {Mima}(1996)}]{sakagami96}%
  \BibitemOpen
  \bibfield  {author} {\bibinfo {author} {\bibfnamefont {H.}~\bibnamefont
  {Sakagami}}\ and\ \bibinfo {author} {\bibfnamefont {K.}~\bibnamefont
  {Mima}},\ }\href {\doibase 10.1103/PhysRevE.54.1870} {\bibfield  {journal}
  {\bibinfo  {journal} {Phys. Rev. E}\ }\textbf {\bibinfo {volume} {54}},\
  \bibinfo {pages} {1870} (\bibinfo {year} {1996})}\BibitemShut {NoStop}%
\end{thebibliography}

%


\end{document}